\documentclass[sigconf]{acmart}
\usepackage{wrapfig}
\usepackage{hyperref}
\usepackage{caption}

\AtBeginDocument{%
  }

\setcopyright{acmlicensed}
\copyrightyear{2025}
\acmYear{2025}
\acmDOI{XXXXXXX.XXXXXXX}

\acmConference[Conference acronym 'XX]{Make sure to enter the correct
  conference title from your rights confirmation email}{June 03--05,
  2024}{Woodstock, NY}
\acmISBN{978-1-4503-XXXX-X/18/06}

\begin{document}

%%
%% The "title" command has an optional parameter,
%% allowing the author to define a "short title" to be used in page headers.
\title{From Division to Unity: A Large-Scale Study on the Emergence of Computational Social Science, 1990–2021}

%%
%% The "author" command and its associated commands are used to define
%% the authors and their affiliations.
%% Of note is the shared affiliation of the first two authors, and the
%% "authornote" and "authornotemark" commands
%% used to denote shared contribution to the research.

\author{Honglin Bao}
\authornote{These authors contributed equally to this research.}
\affiliation{%
  \institution{Knowledge Lab and UChicago Data Science Institute}
  \city{Chicago, IL}
  \country{USA}}
\email{honglinbao@uchicago.edu}

\author{Jiawei Zhang}
\authornotemark[1]
\affiliation{%
  \institution{UChicago Data Science Institute}
 \city{Chicago, IL}
  \country{USA}}
  
\author{Mingxuan Cao}
\authornotemark[1]
\affiliation{%
  \institution{UChicago Data Science Institute}
  \city{Chicago, IL}
  \country{USA}}
%\email{honglinbao@uchicago.edu}

%\affiliation{%
%  \institution{UChicago Data Science Institute}
%  \city{Chicago, IL}
%  \country{USA}}

\author{James A. Evans}
\affiliation{%
  \institution{Knowledge Lab, UChicago Data Science Institute and Sociology, Santa Fe Institute, Google}
  \city{Chicago, IL}
  \country{USA}}
\email{jevans@uchicago.edu}

%%
%% By default, the full list of authors will be used in the page
%% headers. Often, this list is too long, and will overlap
%% other information printed in the page headers. This command allows
%% the author to define a more concise list
%% of authors' names for this purpose.
\renewcommand{\shortauthors}{Bao et al.}

%%
%% The abstract is a short summary of the work to be presented in the
%% article.
\begin{abstract}
We present a comprehensive study on the emergence of Computational Social Science (CSS) — an interdisciplinary field leveraging computational methods to address social science questions — and its impact on adjacent social sciences. We trained a robust CSS classifier using papers from CSS-focused venues and applied it to 11 million papers spanning 1990 to 2021. Our analysis yielded three key findings. \textbf{First}, there were two critical inflections in the rise of CSS. The first occurred around 2005 when psychology, politics, and sociology began engaging with CSS. The second emerged in approximately 2014 when economics finally joined the trend. Sociology is currently the most engaged with CSS. \textbf{Second}, using the density of yearly knowledge embeddings constructed by advanced transformer models, we observed that CSS initially lacked a cohesive identity. From the early 2000s to 2014, however, it began to form a distinct cluster, creating boundaries between CSS and other social sciences, particularly in politics and sociology. After 2014, these boundaries faded, and CSS increasingly blended with the social sciences. \textbf{Third}, shared data-driven methods homogenized CSS papers across disciplines, with politics and economics showing the most alignment due to the combined influence of CSS and causal identification. Nevertheless, non-CSS papers in sociology, psychology, and politics became more divergent. Taken together, these findings highlight the dynamics of division and unity as new disciplines emerge within existing knowledge landscapes. A live demo of CSS evolution can be found in 
\href{https://evolution-css.netlify.app/}{\textcolor{blue}{this URL}}.

\end{abstract}

%%
%% Keywords. The author(s) should pick words that accurately describe
%% the work being presented. Separate the keywords with commas.
%% A "teaser" image appears between the author and affiliation
%% information and the body of the document, and typically spans the
%% page.

%\received{20 February 2007}
%\received[revised]{12 March 2009}
%\received[accepted]{5 June 2009}

%%
%% This command processes the author and affiliation and title
%% information and builds the first part of the formatted document.
\maketitle

\section{Introduction}

Interdisciplinary fields are becoming increasingly vital in driving advancements in scientific paradigms. A quintessential example is Computational Social Science (CSS), which represents a paradigm shift in social science research \cite{kuhn1997structure}. Early discussions of CSS trace back to the 1990s, with a primary focus on social simulation and the study of complex social systems. Researchers utilized computer simulations to replicate and analyze social phenomena, exploring how individual behaviors lead to macro-level social patterns. Between 2000 and 2010, CSS experienced a transition from simulation-based approaches to data-driven methodologies \cite{lazer2009computational}. The proliferation of the internet made large-scale human trace data accessible. Researchers began employing data mining, machine learning, and other computational techniques to analyze real-world social interactions. This shift significantly expanded the depth and breadth of CSS research, enabling the field to address more complex and large-scale social science issues. This progression offers a unique opportunity to study the diffusion of new knowledge across scientific communities and the transformations it generates. Traditionally, research on diffusion has focused on specific concepts or individual products, such as the adoption of a new method within a field \cite{garg2024causal} or the acceptance of a new product \cite{meade2006diffusion, bao2024chatgpt}. Some studies take a broader perspective, examining the history, development, and style of a field \cite{yan2024cultural}. However, the connections between the adoption of new knowledge at scale and the evolution of fields remain underexplored. How does an entirely new discipline like CSS — an organic amalgamation of novel concepts, methods, and research questions — shape the development of adjacent fields’ paradigms and approaches? How are the boundaries between new and existing disciplinary knowledge formed, negotiated, and transformed over time? What mechanisms enable interdisciplinary fields to gain legitimacy and influence? 

Our \textit{\textbf{contributions}} in this paper represent the first systematic attempt, to our knowledge, to quantify the evolution of the impactful interdisciplinary field of CSS at scale and its transformative impact on the social sciences. We address these questions by analyzing a comprehensive dataset of 11 million social science papers spanning 32 years. Using advanced transformer models from natural language processing, we generated yearly embeddings for these papers and developed an ensemble textual classifier to identify CSS-related papers. These efforts advance our understanding of innovation diffusion at a systemic level, moving beyond the adoption of isolated concepts or technologies to encompass the transformation of entire fields of study. CSS, with its dual role as both a unifying and divisive force, serves as a compelling case study for understanding how emerging fields redefine the boundaries of traditional disciplines. %and foster new avenues of inquiry.

\section{Data and Methods}

\subsection{Data}

We collected all 11,527,508 papers and abstracts from Microsoft Academic Graph (MAG) spanning major social sciences during the period of active discussions of CSS, roughly from 1990 to 2021. After 2021 MAG no longer maintained comparably large sizes of papers per year and thus we exclude them. We also exclude non-English papers detected by the widely-used fastText language identifier \cite{joulin2016fasttext}. The final data set includes 10,981,104 papers within the fields of sociology (17.79\%), economics (16.94\%), political science (28.17\%), and psychology (37.09\%). Notably, we did not include papers labeled “communication”, despite recognizing that communication science has been a significant driver of CSS. The reason for this exclusion is that papers’ fields labeled as “communication” make up only a small fraction of social science papers in MAG (less than 2\%) and noisily overlap with telecommunications engineering. Our work relies on the quality of data, and we hope to motivate future studies that could use additional datasets, such as Semantic Scholar,  OpenAlex, and the Web of Science, to complement our findings.

\subsection{Classifier}

We designed a text classifier to identify CSS papers. In the open source community, there is a series of well regarded collections called “awesome + field name” to introduce resources for different fields. We utilized the “Awesome Computational Social Science” list, curated by the Department of Computational Social Science at GESIS – Leibniz Institute for the Social Sciences \footnote{\href{https://github.com/gesiscss/awesome-computational-social-science}{https://github.com/gesiscss/awesome-computational-social-science}}. This list highlights representative CSS conferences and journals, which served as the source for our ground truth 1 labels. We collected papers and abstracts published between 1990 and 2021 from the following journals and conferences: \textit{Big Data \& Society}, \textit{Journal of Social Computing}, \textit{Social Science Computer Review}, \textit{Journal of Computational Social Science}, \textit{International Conference on Computational Social Science}, \textit{Journal of Artificial Societies and Social Simulation}, and \textit{Computational Economics}. From this set, we randomly sampled 2,500 papers to include in the training set as ground truth 1. Some venues listed in the Awesome project were excluded: overly general journals (e.g., \textit{Nature}), overly specialized outlets (e.g., \textit{Complex Networks} and \textit{New Media and Society}), and data science publications not closely tied to social science. To construct 0 labels in the training set, we employed a Word2Vec model \cite{mikolov2013word2vec} to generate an embedding for all abstracts (including titles) in each year with a vector size of 100 and a context window of 5 tokens. We identified the position of the word “computational” in the embedding space and retrieved its 500 nearest neighbors. We then randomly sampled ~20 abstracts per year for each field that did not include any of these 500 keywords (in total 20 $\times$ 32 years $\times$ 4 fields = 2560 abstracts). These abstracts were then manually reviewed to ensure they are not CSS papers. Finally we have a balanced training set across years with 2500 CSS==1 labels, 2500 CSS==0 labels, and their abstracts.

We excluded the newline character $\textbackslash n$ for training and adjusted token weights using the TF-IDF (Term Frequency-Inverse Document Frequency) vectorizer. This approach highlights the importance of unique tokens, such as CSS-specific terms like “algorithmic,” across the corpus. Our classification model is an ensemble of four commonly used text classifiers with two for linear classifiable features and two for non-linear, accommodating the potentially diverse geometries of CSS features: Support Vector Machine (with a linear kernel), Logistic Regression, Random Forest (with 100 estimators), and Gradient Boosting Decision Tree (with 100 estimators). Each classifier applies a distinct logic \cite{bao2024chatgpt}: SVM maximizes decision boundaries, Logistic Regression models the linear relationship between features and labels, Random Forest ranks feature importance, and Gradient Boosting Decision Trees iteratively minimize error. Each classifier outputs a probability of the text being labeled as 1. We calculate the final label by averaging these probabilities: if the average probability exceeds 0.5, the text is labeled as 1; otherwise, it is labeled as 0. The classifier achieves accurate results on an 80\%-20\% training-testing data split:

\begin{table}[h!]
    \centering
    \caption{Summary results of the classifier}
    \resizebox{0.5 \textwidth}{!}{ % Adjust to fit within the width of the page
        \begin{tabular}{|c|c|c|c|c|c|c|}
            \hline
            \textbf{Metric} & Accuracy & F1 & Precision & ROC-AUC &  FP rate & FN rate \\
            \hline
            \textbf{Value} & 0.9769 & 0.9765 & 0.9629 & 0.9958 &  0.0360 & 0.0095 \\
            \hline
        \end{tabular}
    }
    \label{tab:classifier_summary}
\end{table}

\subsection{Embeddings}

We used the SPECTER2 model \cite{singh2023scirepeval} to generate yearly embeddings for papers to analyze the emergence and diffusion of CSS over time. Unlike word-level models such as Word2Vec \cite{mikolov2013word2vec}, which are unsuitable for capturing the overall semantics of abstracts, or general-purpose sentence-level models like Sentence-BERT \cite{reimers2019sentencebert}, which lack domain-specific adaptation, or even scientific-text models like SciBERT \cite{beltagy2019scibert}, which do not integrate citation information, SPECTER2 is pre-trained on large-scale scientific literature. It employs an encoder-only transformer structure and uses self-attention to capture contextual relationships, with citation-based contrastive learning to embed domain-specific semantic links, such as thematic similarity. These capabilities render SPECTER2 exceptionally suited for tasks that require nuanced sentence-level embeddings across scientific disciplines, exhibiting superior performance over most existing representation methods \cite{singh2023scirepeval}. The full text of the title and abstract of a paper was tokenized with a maximum length of 512 tokens covering the vast majority of the length and processed in batches of 32. The [CLS] token embedding from the final transformer layer was used as a pooled representation, encapsulating the semantic meaning at the abstract level. Yearly Word2Vec models \cite{mikolov2013word2vec} were built on titles and abstracts to analyze word relationships.

%we also constructed Word2Vec models \cite{mikolov2013word2vec} for titles and abstracts each year with a vector size of 100 and a context window of 5 tokens.

\section{Results}

\subsection{The reactions of social sciences to CSS}
%before 2000, CSS had limited influence within the social sciences. 
As shown in Figure 1, the prominence of CSS began to grow between 2000 and 2005, first in psychology. However, sociology experienced a notable decline in interest during this period, likely due to skepticism toward social simulation approaches. This skepticism stemmed from the lack of empirical validation, the limited computational capabilities, and the inconsistency issue of the model design \cite{lazer2009computational}. This trajectory shifted after 2005 with the advent of big data and advanced computational tools. These advancements transformed social simulation into a data-driven approach, albeit under the same CSS label, leading to broader acceptance and growth of CSS in sociology and politics after 2005. Economists entered the CSS landscape relatively late, but their engagement saw a dramatic surge after 2014, likely due to the predictive power of machine learning and the rising popularity of AI. CSS in sociology also experienced a significant surge in 2014, accompanied by a smaller increase in psychology. The engagement of sociology with CSS is the highest currently, reaching approximately 8-9\%. Interpretations of these engagements should also consider the low error rates of our classifier (\textasciitilde 1\% false negative and \textasciitilde 3\% false positive rates). 

\begin{wrapfigure}{r}{0.27\textwidth} 
  \centering
  \captionsetup{font=footnotesize} % Adjust caption font size
  \includegraphics[width=\linewidth]{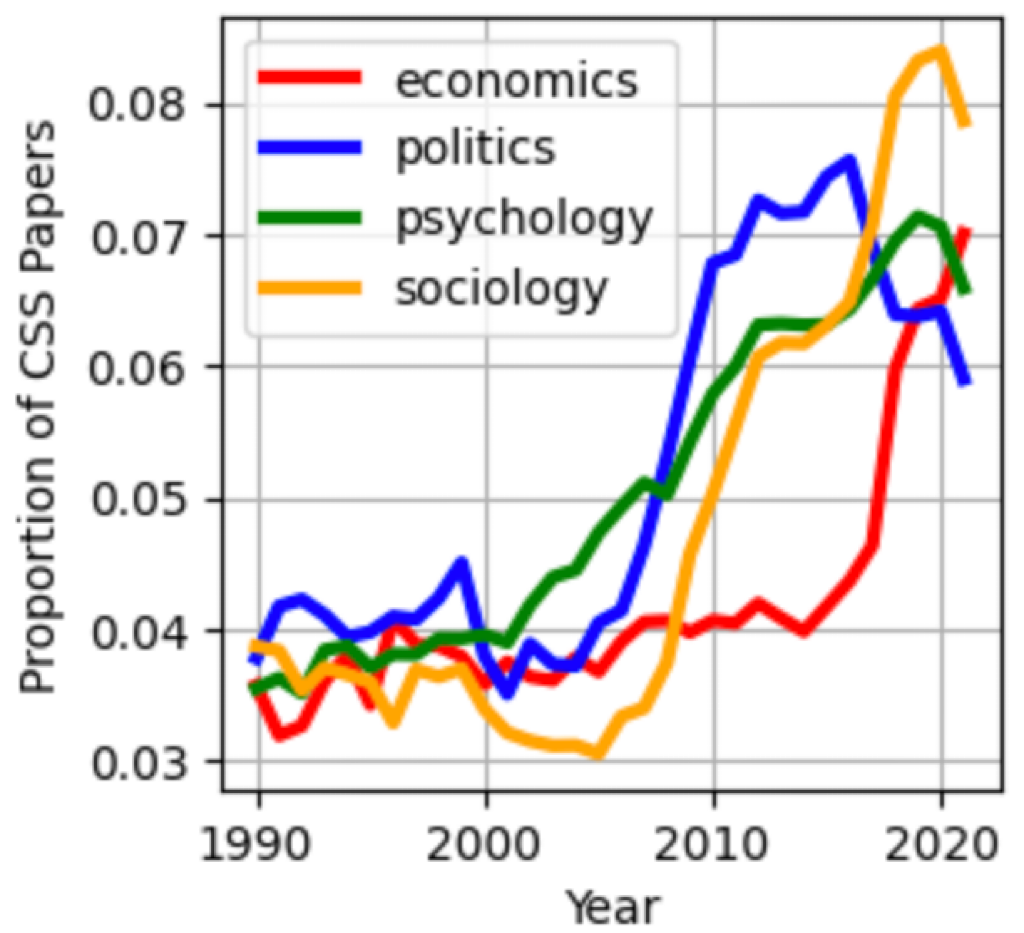}
  \caption{The reactions of social sciences to CSS from 1990-2021.}
\end{wrapfigure}

Interestingly, CSS papers in political science began to decline in the most recent period. One possible explanation is the relationship between political science and the shifting boundary of our sample, with the increasing fusion of political science and communication studies. Many political science papers including CSS ones focus on political communication and social media. Nevertheless, their field labels often overlap, with some being labeled as “communication” excluded from the 32-year sample. In yearly Word2Vec embeddings, we pinpointed 500 words that are close to the midpoint between “politics” and “political” as well as between “communication” and “media.” By bootstrapping their similarity distributions 1000 times, we found that the average similarity rose from 0.68$ \pm $0.01 in 1990 to over 0.9 post-2016. Due to data limitations it remains unclear whether the weak decline in CSS observed in sociology and psychology after 2020 is merely noise or indicative of a continuing trend.

\subsection{Division-unity dynamics of CSS evolution}

\begin{figure}[h]
  \centering
  \captionsetup{font=footnotesize} % Adjust caption font size
  \includegraphics[height=0.19\textheight]{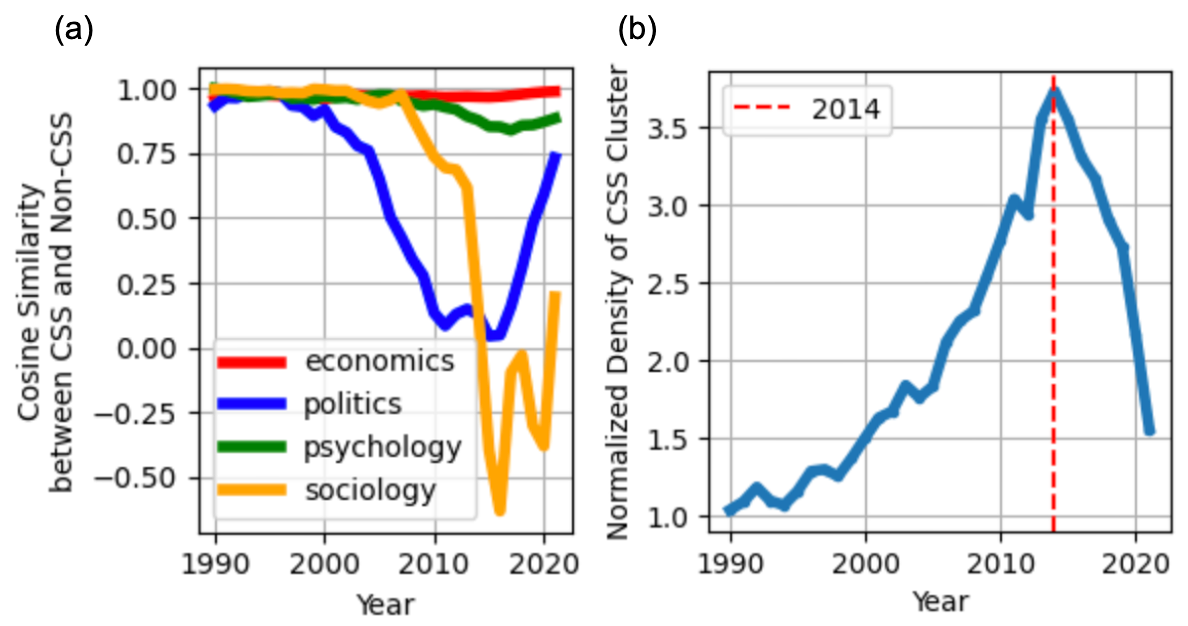}
  \caption{CSS in the embedding space. Panel (a) illustrates the cosine similarity between the central embeddings of CSS papers and non-CSS papers across different years and fields. Panel (b) depicts the dynamics of the normalized density of CSS papers over time.}
\end{figure}

We analyzed the growth and evolution of CSS through its SPECTER2 embedding space. We calculated the similarity between centers in the embedding space of CSS and non-CSS papers within a field across years (Figure 2 panel a). From 1990 to 2000, the similarity between the two was nearly 1. CSS papers were evenly distributed across the knowledge space without a distinct identity or a differentiation from other social sciences (Figure 3). After 2000, however, we observed a notable decline in similarity, followed by a rebound beginning around mid-2010, except for economics. We further calculated the normalized density of CSS papers over time (Figure 2 panel b). For each year, we identified the center of CSS papers in embedding space and randomly sampled 5,000 surrounding papers (nodes) with the highest cosine similarity. We then calculated the proportion of CSS papers in the sampled set, normalized by the overall proportion of CSS papers that year. A normalized density close to 1 would indicate that CSS papers are evenly distributed, whereas an increased value suggests clustering and a decreased value implies diffusion. The density of CSS papers steadily increased before 2014 (Figure 2 panel b), reflecting a gradual clustering that established a distinct identity for CSS papers in embedding space (Figure 3), forming boundaries that differentiated them from others (Figure 2 panel a). Sociology exhibited the most pronounced separation, followed by politics, psychology, and economics. Papers in embedding space form distinct and recognizable clusters, enabling us to approximate clusters using their centers. To confirm these approximations, we retrieved 500 papers surrounding the center and bootstrapped the similarity distributions 1000 times, yielding consistent results. To represent the central position of the majority of papers of a cluster, we used their median.

\begin{wrapfigure}{l}{0.27\textwidth}
  \centering
  \captionsetup{font=footnotesize} % Adjust caption font size
  \includegraphics[width=\linewidth]{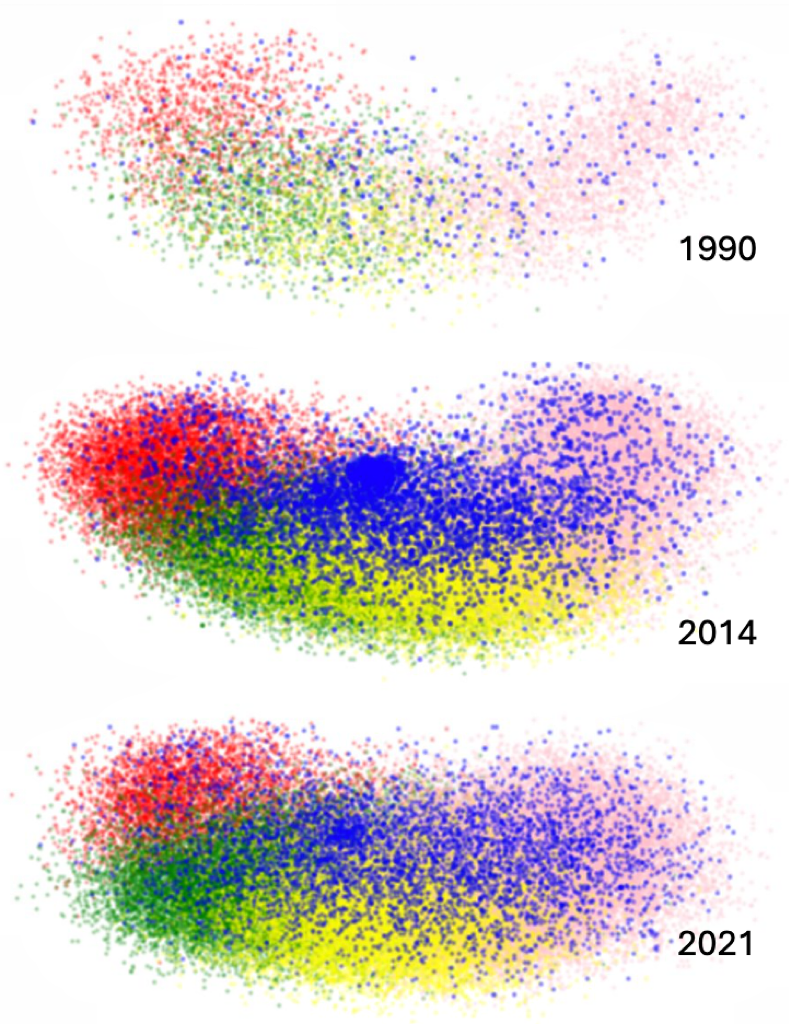}
  \caption{\href{https://evolution-css.netlify.app/}{\textcolor{blue}{Visualization}} of CSS evolution in 1990 (no clustering), 2014 (an identifiable cluster), and 2021 (cluster faded). Principal component analysis is used to reduce dimensions. Economics: red, politics: green, psychology: pink, sociology: yellow, CSS papers: blue. The range of the x and y axes for the plots of these three years is from -6 to 6. To ensure efficient visualization at scale, we sampled 10\% of papers per year for this demo.}
\end{wrapfigure}

Taking sociology as an example, early CSS was often criticized by sociologists for being overly descriptive. 
Elmer likened the early phase of CSS to puberty — a period before mid-2010 in our analysis – when scientific fields often test their boundaries and refine their identity and uniqueness in relation to other disciplines \cite{elmer2023validation}. Nevertheless, after 2014, CSS began to diffuse back into non-CSS domains, evidenced by increased similarity (Figure 2 panel a), declining normalized density (Figure 2 panel b), and fading boundaries (Figure 3). This transition highlights CSS’s broader impact and growing influence across the social sciences, representing the second phase of CSS diffusion. 

%, with many arguing that its acceptance within mainstream social sciences hinges on its capacity to advance theoretical development \cite{edelmann2020css}. 

\subsection{How CSS shapes the social sciences}

Shared computational methods could bring fields closer together. While social sciences may also become increasingly distinct as they strive to maintain their unique identity, particularly non-CSS ones. To test this, we locate the centers of CSS/non-CSS papers in each pair of disciplines, and study if their CSS moves closer than their non-CSS. As shown in Figure 4, computational methods increase the similarity between disciplines, as observed in pairs of economics-politics, economics-psychology, sociology-psychology, and politics-psychology, although in recent developments CSS in psychology appears to be diverging from CSS in sociology and politics. For the economics-sociology and politics-sociology pairs, simulation-based CSS methods (pre-2010) tend to position sociology as more distant from both economics and politics. In contrast, data-driven CSS methods bring sociology closer to these disciplines. Notably, after 2010, CSS clusters in economics and politics almost overlap within the knowledge space, and even non-CSS clusters exhibit an increasingly high similarity. An additional unifying factor that homogenizes them could be causal identification. Using yearly Word2Vec embeddings, we identified 500 words proximal to the midpoint between “politics” and “political” as well as between “causal,” “cause,” and “causality.” By bootstrapping their similarity distributions 1000 times, we observed that the average similarity increased from -0.32$ \pm $0.01 in 1990 to 0.96$ \pm $0.01 in 2021. %Orientations towards data-driven empiricism and causal identification in contemporary scholarship have increasingly blurred the lines between economics and politics.

\begin{wrapfigure}{l}{0.27\textwidth}
  \centering
  \captionsetup{font=footnotesize} % Adjust caption font size
  \includegraphics[width=\linewidth]{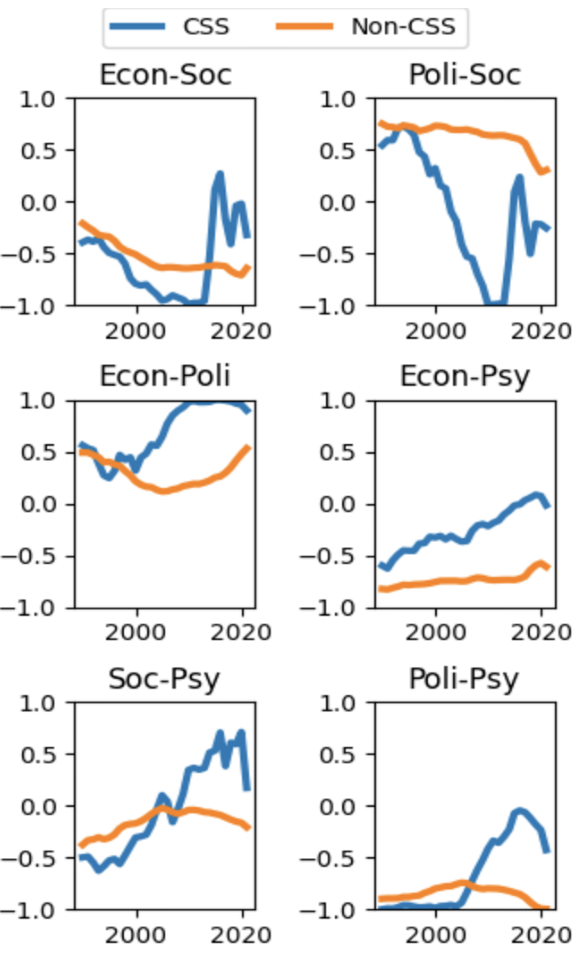}
  \caption{CSS in a pair of disciplines moves closer than their non-CSS. Y-axis: cosine similarity. X-axis: year.}
\end{wrapfigure}

Non-CSS papers exhibit a different pattern. In economics and sociology their relations remained almost unchanged after CSS was established. In economics and psychology they show a tiny increase in similarity, potentially due to their shared experimental methodologies. In sociology and politics, however, the divergence between their non-CSS papers began even before CSS gained widespread adoption, accelerating in the mid-2010s as CSS established a clear disciplinary identity and broadened its influence afterward. In contrast to the growing similarity observed among CSS papers in sociology-psychology and politics-psychology pairs, non-CSS papers underwent a notable shift around 2005 — transitioning from increasing to decreasing similarity — when CSS began to grow and data-driven methods became more prevalent. These trends likely reflect deliberate efforts to preserve the distinctiveness of non-CSS research across fields in response to CSS’s rising prominence.

\section{Conclusions}
%The rise of CSS over the past three decades represents a significant transformation in how social science research is conceptualized and conducted. 

While isolating clean causal effects—such as the precise treatment of CSS's emergence—is challenging, a comprehensive analysis of 11 million papers sheds light on several critical aspects of how CSS evolves and shapes the social sciences. Disciplines vary in how they embrace CSS, reflecting their deep-seated traditions of methods and theories. Fields with strong quantitative traditions, such as political science and psychology, have found it easier to integrate computational approaches. Sociology, with strong qualitative/theoretical traditions, faced more tension reconciling CSS with its existing sociological paradigms. Economics, with its highly mathematical and causality-driven research traditions, responded more slowly to the relatively descriptive, data-focus CSS but ultimately began to embrace these methods rapidly after 2014. The temporal patterns observed in the development of CSS reveal a complex process of identity formation and integration within the broader academic landscape. Two critical inflection points in CSS development are qualitatively distinct: the first, in which economics does not participate, marks CSS as a distinguishable cluster with boundaries separating it from other social sciences, while the second dissolves these boundaries. Efforts towards CSS were initially scattered, as researchers experimented with computational methods independently. Between the early 2000s and 2014, CSS began to cluster, establishing a recognizable identity while creating boundaries that both protected and separated it from mainstream disciplines. After 2014, CSS methods diffused widely, becoming more accepted and integrated. These transitions can homogenize some fields, like economics and political science, while driving non-CSS research further apart from each other.

Our research is not without limitations. The exclusion of the “communication” field due to identification challenges potentially omits a critical area, particularly given the rise of digital media, and acts as a boundary to adjacent disciplines, especially sociology and political science. On the other hand, our research was limited to findings up until 2021. Continuing research may require the design of new classifiers, due to the emergence of ChatGPT in 2022, but there is no doubt that language models and GenAI mark an exciting turning point in CSS. It would also be valuable to explore how top journals in individual disciplines perceive and prioritize CSS over time, how computing fields participate in CSS, as well as the demographic information of authors and topic-level preferences in CSS versus non-CSS papers.

%%
%% The acknowledgments section is defined using the "acks" environment
%% (and NOT an unnumbered section). This ensures the proper
%% identification of the section in the article metadata, and the
%% consistent spelling of the heading.
%\begin{acks}
\smallskip
\textbf{Acknowledgments}: We thank Alex Kale (UChicago Data Science Institute) for his feedback.
%\end{acks}

%%
%% The next two lines define the bibliography style to be used, and
%% the bibliography file.
\bibliographystyle{ACM-Reference-Format}
\bibliography{reference}

\end{document}